\documentclass[showpacs,preprintnumbers,amsmath,amssymb,twocolumn]{revtex4}

\usepackage{graphicx}
\usepackage{dcolumn}
\usepackage{bm}

\begin{document}

\title{On the Ground State of Two Flavor Color Superconductor}
\author{Lianyi He, Meng Jin and Pengfei Zhuang}
\affiliation{Physics Department, Tsinghua University, Beijing
100084, China}

\begin{abstract}
The diquark condensate susceptibility in neutral color
superconductor at moderate baryon density is calculated in the
frame of two flavor Nambu-Jona-Lasinio model. When color chemical
potential is introduced to keep charge neutrality, the diquark
condensate susceptibility is negative in the directions without
diquark condensate in color space, which may be regarded as a
signal of the instability of the conventional ground state with
only diquark condensate in the color $3$ direction.
\end{abstract}

\date{\today}

\maketitle

It is generally accepted that, the cold and dense quark matter
favors the formation of diquark condensate and is in the
superconducting phase\cite{cscreview}, which may be realized in
compact stars and, in very optimistic cases, even in heavy-ion
collisions. To have a stable and macroscopic color superconductor,
one should take into account the electric and color charge
neutrality condition\cite{neutral-1,neutral-2} which leads to a
new phase, the gapless color superconductivity\cite{gapless} or
the breached pairing phase\cite{bp}. In this new phase, the most
probable temperature is finite but not zero\cite{calabash}, and
the Meissner screening masses squared can be
negative\cite{meissner}. In two flavor case, the color neutrality
can be satisfied by introducing a color chemical potential $\mu_8$
in the four-fermion interaction theory at moderate baryon
density\cite{neutral-2}, or by a dynamic generation of a
condensation of gluon field $A_0^8$ in the frame of perturbative
QCD at extremely high baryon density\cite{neutral-3} where the
back ground gluon field $\langle A_0^8\rangle$ plays the role of
the color chemical potential $\mu_8$.

In conventional approach to investigating two flavor color
superconductivity, one chooses the first two colors participating
in the diquark condensate and let the third one be
free\cite{schafer}. When the Lagrangian in the study is color
$SU(3)$ symmetric, the ground states with different color breaking
directions can be connected by some $SU(3)$ transformations in
color space and hence considering only one direction is enough to
describe the system. However, when the color-charge neutrality is
taken into account, one has to introduce nonzero color chemical
potentials at moderate baryon density in the
Nambu--Jona-Lasinio(NJL) type model, and the global color $SU(3)$
symmetry of the NJL model is then explicitly broken. In this case
the state with only two colors participating in the Cooper pairing
is probably not the ground state, and one has to calculate
thermodynamical potentials for different states and chooses the
one with the minimum thermodynamical potential as the ground
state. In this paper, we will calculate the diquark condensate
susceptibility in the other two directions without diquark
condensate in color space, from which we can judge if the
conventional ground state is stable when the color charge
neutrality in considered.

We work in the widely used NJL model applied to
quarks\cite{njlquark}. The model has been successfully used to
study chiral symmetry restoration\cite{njlquark,njlchiral},
isospin symmetry spontaneously broken\cite{barducci,he-1}, and
color
superconductivity\cite{neutral-2,gapless,calabash,schwarz,zhuang,ebert,ruster,he-2,huang}
at moderate baryon density. In chiral limit, the flavor SU(2)
Lagrangian density including quark-quark interaction sector is
defined as
\begin{eqnarray}
\label{njl} {\cal L} &=&
\bar{\psi}\left(i\gamma^{\mu}\partial_{\mu}+\mu\gamma_0\right)\psi
+G_S\left(\left(\bar{\psi}\psi\right)^2+\left(\bar{\psi}i\gamma_5\vec{
\tau}\psi\right)^2
\right)\nonumber\\
&+&G_D\left(\bar\psi^c_{i\alpha}
i\gamma^5\epsilon^{ij}\epsilon^{\alpha\beta\gamma}\psi_{j\beta}\right)\left(\bar\psi_{i\alpha}
i\gamma^5\epsilon^{ij}\epsilon^{\alpha\beta\gamma}\psi^c_{j\beta}\right)
\ ,
\end{eqnarray}
where $G_S$ and $G_D$ are, respectively, coupling constants in
color singlet and anti-triplet channels, $\psi^c = C\bar\psi^T$
and $\bar\psi^c = \psi^T C$ are charge-conjugate spinors, $C =
i\gamma^2\gamma^0$ is the charge conjugation matrix, the quark
field $\psi_{i\alpha}$ with flavor index $i$ and color index
$\alpha$ is a flavor doublet and color triplet as well as a
four-component Dirac spinor, ${\bf \tau} =(\tau_1, \tau_2,
\tau_3)$ are Pauli matrices in flavor space, and $\epsilon^{ij}$
and $\epsilon^{\alpha\beta\gamma}$ are, respectively, totally
antisymmetric tensors in flavor and color spaces. We focus in the
following on the color symmetry breaking phase with nonzero
diquark condensates defined as
\begin{equation}
\label{delta} \Delta_\gamma = -2G_D\langle
\bar\psi^c_{i\alpha}i\gamma^5\epsilon^{ij}\epsilon^{\alpha\beta\gamma}\psi_{j\beta}\rangle,
\ \ \ \ \ \gamma=1,2,3\ .
\end{equation}
To simplify the calculation, we assume that the chiral symmetry is
already restored in this phase. This assumption is confirmed when
the coupling constant $G_D$ in the diquark channel is not too
large\cite{zhuang}. To ensure color and electric neutralities, one
should introduce a set of color chemical potentials
${\mu_a}(a=1,2,...,8)$ with respect to color charges
$Q_1,Q_2,...,Q_8$ and an electric chemical potential $\mu_e$ with
respect to the electric charge (electrons are included). The
ground state of the system is determined by the gap equations
which can be obtained by minimizing the thermodynamical potential
$\Omega$,
\begin{equation}
\label{gap1}
\partial\Omega/\partial\Delta_\gamma=0\ ,
\end{equation}
and the charge neutrality condition,
\begin{eqnarray}
\label{neutrality1} Q_e&=&-\partial\Omega/\partial\mu_e=0
,\nonumber\\ Q_a&=&-\partial\Omega/\partial\mu_a=0,
~~(a=1,\cdots,8).
\end{eqnarray}
The above twelve coupled equations (\ref{gap1}) and
(\ref{neutrality1}) determine self-consistently the physical
condensates $\Delta_\gamma$ and the chemical potentials $\mu_e$
and $\mu_i$ as functions of temperature $T$ and baryon chemical
potential $\mu_b$.

We first consider the conventional ground state with
\begin{equation}
\label{conventional} \Delta_1=\Delta_2=0\ ,\ \
\Delta_3\equiv\Delta\neq 0\ .
\end{equation}
In this case, $Q_1,\cdots,Q_7$ vanishes automatically, and only
$Q_8\not=0$. Therefore, we can only introduce the color chemical
potential $\mu_8$ with respect to the color charge $Q_8$ to ensure
color neutrality, and the quark chemical potential matrix
\begin{equation}
\label{mu1}
\mu=diag(\mu_{u1},\mu_{u2},\mu_{u3},\mu_{d1},\mu_{d2},\mu_{d3})
\end{equation}
in color and flavor space can be expressed in terms of the baryon
chemical potential $\mu_b$, electrical chemical potential $\mu_e$,
and color chemical potentials $\mu_8$,
\begin{eqnarray}
\label{mu2}
&&\mu_{u1}=\mu_{u2}=\mu_b/3-2\mu_e/3+\mu_8/3\ ,\nonumber\\
&&\mu_{u3}=\mu_b/3-2\mu_e/3-2\mu_8/3\ ,\nonumber\\
&&\mu_{d1}=\mu_{d2}=\mu_b/3+\mu_e/3+\mu_8/3\ ,\nonumber\\
&&\mu_{d3}=\mu_b/3+\mu_e/3-2\mu_8/3\ .
\end{eqnarray}
Generally, $\mu_8$ is nonzero, and the color symmetry of the NJL
Lagrangian is explicitly broken down to $SU(2)\bigotimes U(1)$
with generators $T_1, T_2, T_3$ and $T_8$. In mean field
approximation the thermodynamical potential $\Omega_0$ of the
system can be evaluated as
\begin{eqnarray}
\label{omega1} \Omega_0 ={\Delta^2\over 4G_D}-{T\over
2}\sum_n\int{d^3 {\bf p}\over (2\pi)^3}Tr \ln {\cal
S}^{-1}_{mf}+{\mu_e^4\over 12\pi^2}\ ,
\end{eqnarray}
where the last term is the contribution from the electron gas. In
the modified 12-dimensional Nambu-Gorkov space with color and
flavor indices, defined by the field vector
\begin{eqnarray}
\label{gokov} \bar\Psi&=&(\bar\psi_{u1},\ \bar\psi^c_{d2},\ \bar\psi_{d2},\ \bar\psi^c_{u1},\ \bar\psi_{d1},\ \bar\psi^c_{u2}, \nonumber\\
&&\bar\psi_{u2},\ \bar\psi^c_{d1},\ \bar\psi_{u3},\
\bar\psi^c_{u3},\ \bar\psi_{d3},\ \bar\psi^c_{d3})\ ,
\end{eqnarray}
the mean field quark propagator ${\cal S}^{-1}_{mf}$ is diagonal
and can be expressed as
\begin{equation}
\label{smf1} {\cal
S}_{mf}^{-1}=diag\left(\begin{array}{cccccccccccc}{\cal
S}_{u1,d2}^{-\Delta}&{\cal S}_{d2,u1}^{-\Delta}&{\cal
S}_{d1,u2}^{\Delta}&{\cal S}_{u2,d1}^{\Delta}&{\cal
S}_{u3,u3}^0&{\cal S}_{d3,d3}^0
\end{array}\right)
\end{equation}
where the diagonal blocks are defined as
\begin{equation}
{\cal S}_{i\alpha,j\beta}^\chi=\left(\begin{array}{cc}[{\cal
G}_0^+]^{-1}_{i\alpha}&i\gamma_5\chi\\i\gamma_5\chi&[{\cal
G}_0^-]^{-1}_{j\beta}
\end{array}\right)
\end{equation}
with the free quark propagators
\begin{equation}
\label{g0} \left[{\cal
G}_0^\pm\right]_{i\alpha}^{-1}=\left(i\omega_n\gamma_0-{\bf
p}\cdot{\bf \gamma}\pm\mu_{i\alpha}\gamma_0\right)\ .
\end{equation}

Since the quark propagator ${\cal S}^{-1}_{mf}$ is diagonal, we
can analytically take its trace in the color, flavor and Dirac
spaces and make the Matsubara frequency summation, the
thermodynamic potential can be expressed as an explicit function
of $\Delta, \mu_e, \mu_8, \mu_b$ and $T$. Minimizing the known
$\Omega_0$, we then obtain\cite{neutral-2} in the color symmetry
spontaneously breaking phase the gap equation
\begin{equation}
\label{gap2} 1-8G_D\int{d^3{\bf p}\over
(2\pi)^3}\sum_{\epsilon=\pm}{1-f(E^\epsilon_+)-f(E^\epsilon_-)\over
E^\epsilon_\Delta} = 0
\end{equation}
and the charge neutrality condition
\begin{eqnarray}
\label{neutrality2} Q_8 &=& \int{d^3{\bf p}\over
(2\pi)^3}\sum_{\epsilon=\pm}\epsilon\Big[{E^\epsilon_0\over
E^\epsilon_\Delta}\left(1-f(E^\epsilon_+)-f(E^\epsilon_-)\right)\nonumber\\
&+&\left(f(E_{u3}^\epsilon)+f(E_{d3}^\epsilon)\right)\Big]=0\
,\nonumber\\
Q_e &=& \int{d^3{\bf p}\over
(2\pi)^3}\sum_{\epsilon=\pm}\Big[\epsilon{E^\epsilon_0\over
E^\epsilon_\Delta}\left(1-f(E^\epsilon_+)-f(E^\epsilon_-)\right)\nonumber\\
&+&3\left(f(E^\epsilon_+)-f(E^\epsilon_-)\right)-\epsilon
\left(2f(E_{u3}^\epsilon)-f(E_{d3}^\epsilon)\right)\Big]\nonumber\\
&-&{\mu_e^3\over 3\pi^2}=0\ ,
\end{eqnarray}
where the quasi-particle energies are defined as $E^\pm_\mp =
E^\pm_\Delta \mp\delta\mu, E^\pm_\Delta = \sqrt{(|{\bf
p}|\pm\bar\mu)^2+\Delta^2}, E_0^\pm=|{\bf p}|\pm\bar \mu,
E_{u3}^\pm = |{\bf p}|\pm\mu_{u3}$ and $E_{d3}^\pm = |{\bf
p}|\pm\mu_{d3}$ with the two effective chemical potentials
$\bar\mu$ and $\delta\mu$ given by $\bar\mu =
\mu_b/3-\mu_e/6+\mu_8/3$ and $\delta\mu=\mu_e/2$, and
$f(x)=1/\left(e^{x/T}+1\right)$ is the Fermi-Dirac distribution
function. The equations (\ref{gap2}) and (\ref{neutrality2})
determine simultaneously the order parameter $\Delta$ and chemical
potentials $\mu_e$ and $\mu_8$ in the conventional ground state of
neutral color superconductor.

Now we come to the question whether the conventional ground state
defined through (\ref{conventional}) is stable. To answer this
question we calculate the second order derivation of the
thermodynamical potential $\Omega$ with respect to the diquark
condensates $\Delta_1$ and $\Delta_2$,
\begin{equation}
\label{kappa3}\kappa=\frac{\partial^2 \Omega}{\partial
\Delta_1^2}\Big|_{\Delta_1=\Delta_2=0}= \frac{\partial^2
\Omega}{\partial\Delta_2^2}\Big|_{\Delta_1=\Delta_2=0}.
\end{equation}
We call the quantity $\kappa$ diquark condensate susceptibility.

Since the explicit form of $\Omega$ with finite condensates
$\Delta_1$ and $\Delta_2$ is not easy to obtained, we shall use
the method of perturbation. We take
\begin{eqnarray}
\label{perturbation}
\Delta_1&=&\delta_1\ll\Delta\ ,\nonumber\\
\Delta_2&=&\delta_2\ll\Delta\ ,
\end{eqnarray}
and apply the Taylor expansion
\begin{equation}
\label{omega4} \Omega_\delta-\Omega_0={1\over
2}\kappa\left(\delta_1^2+\delta_2^2\right)+...\ .
\end{equation}
Here the perturbed thermodynamical potential $\Omega_\delta$ can
be written as
\begin{eqnarray}
\label{omega3} \Omega_\delta =&&
{\Delta^2+\delta_1^2+\delta_2^2\over 4G_D}+{\mu_e^4\over
12\pi^2}\nonumber\\
&&-{T\over 2}\sum_n\int{d^3{\bf p}\over (2\pi)^3}Tr\ln \left[{\cal
S}^{-1}_{mf}+\Gamma\left(\delta_1,\delta_2\right)\right]
\end{eqnarray}
with the matrix $\Gamma\left(\delta_1,\delta_2\right)$ defined as
\begin{widetext}
\begin{equation}
\label{smf1} \Gamma\left(\delta_1,\delta_2\right)=
\left(\begin{array}{cccccccccccc}
0&0& 0& 0& 0& 0& 0& 0& 0& 0& 0& \delta_2\\
0&0& 0& 0& 0& 0& 0& 0& \delta_1& 0& 0& 0\\
0& 0& 0&0& 0& 0& 0& 0& 0& \delta_1& 0& 0\\
0& 0&0&0 & 0& 0& 0& 0& 0& 0& \delta_2& 0\\
0& 0& 0& 0&0& 0& 0& 0& 0& -\delta_2& 0& 0\\
0& 0& 0& 0&0&0 & 0& 0& 0& 0& -\delta_1& 0\\
0& 0& 0& 0& 0& 0&0 &0 & 0& 0& 0& -\delta_1\\
0& 0& 0& 0& 0& 0& 0& 0& -\delta_2& 0& 0& 0\\
0& \delta_1& 0& 0& 0& 0& 0& -\delta_2& 0& 0& 0& 0\\
0& 0& \delta_1& 0& -\delta_2& 0& 0& 0& 0& 0& 0& 0\\
0& 0& 0& \delta_2& 0& -\delta_1& 0& 0& 0& 0& 0& 0\\
\delta_2& 0& 0& 0& 0& 0& -\delta_1& 0& 0& 0& 0& 0
\end{array}\right).
\end{equation}
\end{widetext}
Taking into account the expansion
\begin{eqnarray}
Tr\ln(A^{-1}+B)=Tr\ln A^{-1}-\sum_{n=1}^\infty\frac{(-1)^n}{n} Tr
(AB)^n
\end{eqnarray}
for general matrices A and B and the relation
\begin{equation}
\label{fact} Tr\left[{\cal
S}_{mf}\Gamma\left(\delta_1,\delta_2\right)\right] = 0\ ,
\end{equation}
which ensures the linear term to be zero, we can expand
$\Omega_\delta$ to the quadratic terms in $\delta_1$ and
$\delta_2$, and obtain the diquark condensate susceptibility
\begin{eqnarray}
\label{kappa1} \kappa = {1\over 2G_D}+{T\over 4}\sum_n\int{d^3{\bf
p}\over (2\pi)^3} Tr\left({\cal
S}_{mf}\Gamma\left(1,0\right)\right)^2\ .
\end{eqnarray}

The next task is to calculate $\kappa$ based on the known $\Delta,
\mu_e$ and $\mu_8$ in the conventional ground state.  After a
somewhat complicated but straightforward algebra calculation and
using the gap equation (\ref{gap2}), $\kappa$ can be simplified as
\begin{equation}
\label{kappa2} \kappa=-2\mu_8 K(0)
\end{equation}
with the function $K(x)$ given by
\begin{eqnarray}
\label{k-1} K(x)&=& \int{d^3{\bf p}\over (2\pi)^3}
\sum_{\epsilon=\pm}\Big[\left({1\over F_1^\epsilon}-{1\over
F_2^\epsilon}\right){f(E_{u3}^\epsilon)+f(E_{d3}^\epsilon)-1\over
E^\epsilon_\Delta}\nonumber\\
&+&\left({1\over F_1^\epsilon} +{1\over
F_2^\epsilon}\right){f(E_+^\epsilon)+f(E_-^\epsilon)-1\over
E_\Delta^\epsilon}\Big]\ ,
\end{eqnarray}
where the $x$-dependence is hidden in $F_{1,2}^\pm$ defined as
\begin{eqnarray}
F_1^\pm (x,|{\bf p}|) &=& x+\mu_8\mp E^\pm_0\mp E^\pm_\Delta \
,\nonumber\\
F_2^\pm (x,|{\bf p}|) &=& x+\mu_8\mp E^\pm_0\pm E^\pm_\Delta \ .
\end{eqnarray}

It is clear that, in the case without considering the color
chemical potential, namely $\mu_8 =0$, we have $\kappa=0$, the
conventional ground state is stable under the perturbation. In
fact, when $\mu_8=0$ the thermodynamic potential depends only on
the quantity $\sqrt{\Delta_1^2+\Delta_2^2+\Delta_3^2}$, thus one
can choose $\Delta_1=\Delta_2=0, \Delta_3\neq0$ without losing
generality. When explicitly considering $\mu_8\neq 0$, the
function $K(x)$ is related to the color charge density $Q_8$
through\cite{he-2}
\begin{equation}
\label{neutrality-3} K(-\mu_8)=2{Q_8\over \Delta^2}=0\ ,
\end{equation}
we can then expand $K(x)$ around $x=-\mu_8$. Since the magnitude
of $\mu_8$ is only a few MeV in the whole color breaking
phase\cite{neutral-2,he-2}, which is much less than the order
parameter $\Delta$, the baryon chemical potential $\mu_b$, and the
momentum cutoff $\Lambda$, we can keep only the linear term in the
expansion,
\begin{eqnarray}
\label{k-2} \kappa &=& -2\mu_8^2K'(-\mu_8)\nonumber\\
&=& -{4\mu_8^2\over \Delta^4}\int{d^3{\bf p}\over
(2\pi)^3}\sum_{\epsilon=\pm}\Big[{\left(E_0^\epsilon\right)^2+\left(E_\Delta^\epsilon\right)^2\over
E_\Delta^\epsilon}\nonumber\\ &&\left(1-f(E_+^\epsilon)-f(E_-^\epsilon)\right)-2E_0^\epsilon \left(1-f(E_{u3}^\epsilon)-f(E_{d3}^\epsilon)\right)\Big]\nonumber\\
&<& -{4\mu_8^2\over \Delta^4}\int{d^3{\bf p}\over
(2\pi)^3}\sum_{\epsilon=\pm}\Big[{\left(E_\Delta^\epsilon-\left(E_{3u}^\epsilon+E_{3d}^\epsilon\right)/2\right)^2\over
E_\Delta^\epsilon}\nonumber\\ &&\left(1-f(E_+^\epsilon)-f(E_-^\epsilon)\right)\Big]\nonumber\\
&<&0\ .
\end{eqnarray}

\begin{figure}
\centering \includegraphics[width=3.5in]{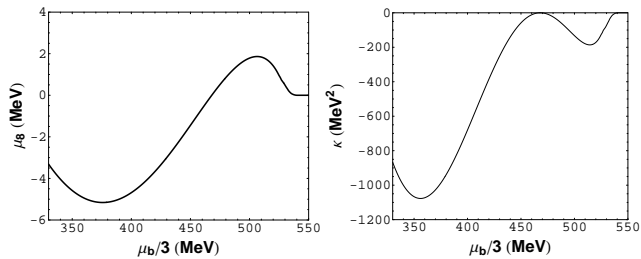}%
\caption{The color chemical potential $\mu_8$ and $\kappa$ defined
in (\ref{kappa2}) as functions of baryon chemical potential
$\mu_b$ at temperature $T=0$ for color superconductor with both
constraints $Q_8=0$ and $Q_e=0$. } \label{fig1}
\end{figure}

\begin{figure}
\centering \includegraphics[width=3.5in]{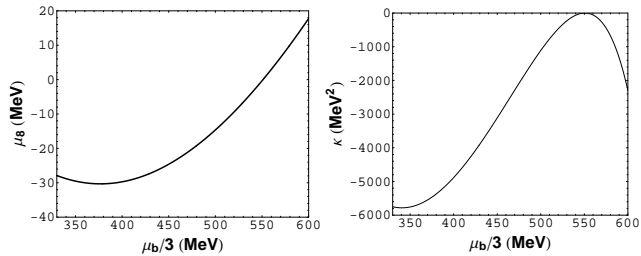}%
\caption{The color chemical potential $\mu_8$ and $\kappa$ defined
in (\ref{kappa2}) as functions of baryon chemical potential
$\mu_b$ at temperature $T=0$ for color superconductor with only
the constraint $Q_8=0$. } \label{fig2}
\end{figure}

Therefore, we have proven analytically that, under the condition
$|\mu_8|\ll \Delta, \mu_b, \Lambda$, $\kappa$ is negative in the
conventional ground state when the color charge neutrality is
considered. This conclusion can also be proven numerically. To do
numerical calculation, we choose the parameters
$G_S=5.01GeV^{-2}$, $\Lambda=0.653GeV$ and $G_D=3G_S/4$ as in
Ref.\cite{neutral-2}. With these parameters the minimum baryon
chemical potential where the color superconductivity phase starts
is $\mu_b/3 = 330$ MeV. The baryon chemical potential dependence
of $\mu_8$ and $\kappa$, calculated through (\ref{gap2}),
(\ref{neutrality2}) and (\ref{kappa2}), is shown in Fig.\ref{fig1}
in the case with both color and electrical chemical potentials and
in Fig.\ref{fig2} with only color chemical potential (namely
taking $\mu_e =0$). We see that, $\kappa$ is negative in the whole
color breaking phase, independent of whether the electrical charge
neutrality is taken into account or not. In fact, in both cases
with and without electrical charge neutrality, $\mu_8$ is really
very small, compared with $\Delta, \mu_b, \Lambda$ which are all
the order of hundreds MeV, the assumption used in the analytic
calculation is safe.


In summary, we have investigated the relation between the explicit
color symmetry breaking and the ground state which reflects the
spontaneous color symmetry breaking. For color superconductor
without color neutrality, the Lagrangian of the system is color
SU(3) symmetrical, and the spontaneous breaking can happen in only
one of the three directions in color space, namely from SU(3) to
SU(2). However, when the color chemical potential is taken into
account, the Lagrangian of NJL model loses its color SU(3)
symmetry explicitly, and therefore the spontaneous color breaking
in only one direction ($\Delta_3$) may be impossible. Through
calculating the diquark condensate susceptibility $\kappa$ in the
other two directions, we have proven analytically and numerically
that $\kappa$ is always negative in the whole color breaking phase
in the frame of flavor SU(2) NJL model. This may be a signal that
the conventional ground state in unstable. More detail
investigations in this aspect will be done in our future works.

{\bf Acknowledgement:} The work is supported by the grants
NSFC10428510, 10435080 10575058 and SRFDP20040003103.

\end{document}